# Structural investigation of uniform ensembles of self-catalyzed GaAs nanowires fabricated by a lithography-free technique

Eero Koivusalo*, Teemu Hakkarainen, Mircea Guina

Optoelectronics Research Centre, Tampere University of Technology, Korkeakoulunkatu 3, FI-33720 Tampere, Finland

*eero.koivusalo@tut.fi

**Abstract**

Structural analysis of self-catalyzed GaAs nanowires (NWs) grown on lithography-free oxide patterns is described with insight on their growth kinetics. Statistical analysis of templates and NWs in different phases of the growth reveals extremely high dimensional uniformity due to a combination of uniform nucleation sites, lack of secondary nucleation of NWs, and self-regulated growth under the effect of nucleation antibunching. Consequently, we observed the first evidence of sub-Poissonian GaAs NW length distributions. The high phase purity of the NWs is demonstrated using complementary transmission electron microscopy (TEM) and high-resolution x-ray diffractometry (HR-XRD). It is also shown that, while NWs are to large extent defect-free with up to 2 μm long twin-free zincblende segments, low temperature micro-photoluminescence spectroscopy reveals that the proportion of structurally disordered sections can be detected from their spectral properties.

**Introduction**

Semiconductor nanowires (NWs) offer versatile possibilities for realization of future electronic and photonic devices. For example, owing to their small footprint one-dimensional NWs enable integration of highly lattice- or thermally- mismatched materials on dissimilar substrates [1]. Combination of the excellent optical and electronic properties of III-V semiconductors and the low cost Si platform is technologically very tempting. The development of NW based technologies has already led to integration of III-V NW based lasers [2,3] and light emitting diodes [4,5] on Si. Furthermore, the large absorption cross-section of NWs can be exploited by NW solar cells [6,7] or NWs can be used as building blocks for high mobility transistors [8,9] and single-photon emitters [10,11]. To this end, it is important to minimize the crystal defects such as polytypism and stacking faults which could affect the device performance. For example, zincblende (ZB) – wurtzite (WZ) interfaces in GaAs NWs have type II band alignment which causes carrier localization along the NW axis and localized recombination at the interfaces [12,13]. Many applications also require homogeneous NW arrays with narrow size distributions as the NW dimensions significantly affect their optical properties [14,15] and cause difficulties in device fabrication.

Self-catalyzed NWs are often grown via self-assembled method on porous $SiO_x$ or native oxide covered wafers. Disadvantages of this method include coupling of the NW nucleation and growth conditions as well as variations in nucleation time for different NWs, which significantly broadens their size distribution [16,17]. Furthermore, the wafer-to-wafer variations and nonuniformity of the oxide properties lead to reproducibility issues in the NW growth [18]. Homogeneous NW arrays may be obtained by selective area growth in an array of holes in $SiO_x$-layer formed using for instance nanoimprint [19] or electron beam lithography [14,20]. However, lithographical techniques require rather complex substrate preparation procedure while exposing it to various chemicals [19]. To mitigate these issues, we have recently introduced a lithography-free droplet-epitaxy-based method for making $Si/SiO_x$ patterns that allow self-catalyzed growth of highly uniform GaAs NWs [21]. In this letter, we provide detailed insight of the growth process of these NWs based on statistical analysis of feature sizes at different stages of the template fabrication and NW growth. The NW microstructure is then assessed by complementary high-resolution transmission electron microscopy (HR-TEM) and high-resolution x-ray diffraction (HR-XRD) techniques. Furthermore, we investigate the correlation between the structural defects and optical properties and show that the proportion of structurally disordered sections in mostly defect-free NWs increases the fine structure in the low temperature micro-photoluminescence spectra.

**Experimental**

Self-catalyzed GaAs NWs were grown on lithography-free oxide patterns using molecular beam epitaxy (MBE). The growth method initiated with droplet epitaxy (DE) of GaAs nanocrystals on oxide-free p-type Si(111) substrates. The native oxide covering the substrates was removed by a dip in aqueous solution of hydrofluoric acid. After the oxide removal, the substrates were immediately transferred to the MBE-reactor and degassed at 640 °C prior to Ga droplet formation at 545 °C by deposition of 0.7 monolayers (ML) of Ga with 0.1 ML/s growth rate calibrated for planar GaAs growth on GaAs(001). The droplets were then crystallized into GaAs by exposing them to $As_2$-flux for 10 min. Subsequently, the templates were removed from the MBE chamber and exposed to air for 18 hours in order to form a thin oxide layer on the Si surface. During this time the nanocrystal density and diameter were characterized by scanning electron microscopy (SEM). The NWs were grown in the second growth step which started with a 30 min thermal annealing at 655 °C for desorbing the GaAs nanocrystals, thus forming a template composed of oxide-free holes on otherwise $SiO_x$-covered Si(111)-surface, which act as nucleation sites for the subsequent NW growth. Further details of the template properties and fabrication process can be found in [21]. A 60 s Ga pre-deposition at 640 °C with Ga flux corresponding to 0.3 MLs$^{-1}$ was performed prior to the NW growth. The NW growth was then initiated in the same substrate temperature by providing an $As_2$ flux with V/III ratio of 9. Four samples were grown with NW growth durations 60, 40, 20 and 5 min, referred as NW1, NW2, NW3, and NW4, respectively. The GaAs nanocrystal densities of the templates used for NW growth were $2 \times 10^8$ cm$^{-2}$. In sample NW1 the growth was terminated by shutting the $As_2$ and Ga fluxes simultaneously, thus preserving the Ga catalyst droplets, whereas samples NW2, NW3, and NW4 were exposed to $As_2$-flux during the sample cool-down thus crystallizing the Ga catalyst droplets.

The crystal structure of the NWs in samples NW1 and NW2 was studied using HR-TEM and HR-XRD. Samples NW1 and NW2 were chosen for characterization in order to see the differences between NWs with preserved (NW1) and crystallized (NW2) Ga catalyst droplets.

A photoluminescence (PL) study was conducted in order to see the effect of Ga droplet crystallization and NW size on the optical properties of the NWs. Power dependent micro-PL measurements were conducted at 10 K temperature using as-grown samples. A 640 nm diode laser with excitation power densities ranging from 0.1 to 960 $W/cm^2$ was used. The excitation beam spot size was 20 µm meaning that several NWs were excited simultaneously. The PL signal was collected using a 1024x256 pixel CCD detector through a spectrometer. Spatial resolution in one of the lateral directions was obtained by defining pixel rows as individual tracks, while for the other lateral direction it was given by the slit placed in front of the spectrograph. The resulting spatial resolution was around 500 nm which allowed us to collect PL signal corresponding to an individual NW or few NWs.

**Results and discussion**

SEM images of samples NW1-NW4 are shown in Fig. 1. Fig. 1(a) represents sample NW1 that exhibits preserved Ga catalyst droplets; Fig 1(b)-(d) show samples NW2-NW4, respectively, which have crystallized droplets. The most prominent feature seen in the SEM images is the size uniformity of the NW arrays. This characteristic was further studied analyzing minimum of 20 NWs from each sample. The length and diameter data of the NWs including the standard deviations is shown in Fig. 2. The growth of NW diameter and length is linear and the NWs do not exhibit tapered morphology despite of the finite radial growth. The standard deviations of the NW length and diameter in sample NW1 are 1.2 % and 3.8 %, respectively. These distributions are similar [19] or even narrower [22] than the ones obtained via selective area growth. It is noteworthy that the standard deviation of the NW length is around 40 nm for all the samples. Hence, the differences in the NW lengths are formed at very early stage of the NW growth and can be attributed to differences in reaching the supersaturation condition of different size Ga droplets [23].

In order to further investigate the evolution of NWs throughout the growth process, we determined the statistics for GaAs nanocrystal diameter in the template material (Fig. 3(a)), Ga catalyst droplet diameter after 60 s Ga pre-deposition (Fig. 3(b)), as well as NW and catalyst droplet diameters and NW length of sample NW1 with 60 min growth duration (Fig. 3(c) and (d)), respectively. The catalyst droplet diameters after Ga pre-deposition (Fig. 3(b)) were determined from an additional sample with only pre-deposited Ga droplets, which were grown on similar template as the NW samples. The densities of GaAs nanocrystal and pre-deposited Ga droplets were determined and found to be equal, which indicates that a Ga droplet is present in all holes in the oxide layer. The standard deviation of the diameter increases in the very beginning of the growth process from 2.3 nm of the pre-deposited Ga droplets (Fig. 3(b)) to 3.8 nm of NWs grown for 5 min (NW4, Fig. 2). On the other hand, it can be immediately seen from Fig. 3 that the size distribution remains remarkably narrow throughout the growth process. The diameter of the GaAs nanocrystals is 51 nm +/-4.9 nm (Fig.3(a)). In the beginning of the second growth step they are thermally desorbed and replaced with Ga catalyst droplets having diameter of 41 nm +/-2.3 nm (Fig. 3(b)), indicating reduction of the diameter distribution between these two stages. This is most likely caused by slight asymmetries observed in the shape of the nanocrystals. It should be also noted that the uncertainty in assessing the feature sizes from SEM pictures is approximately 1-2 nm. In Fig. 3(c) and (d), the distributions of NW diameter and length after 60 min growth are compared to Poissonian distributions

$$F(D, \langle D \rangle) \propto \exp\left[-\frac{(D-\langle D \rangle)^2}{2w\langle D \rangle}\right], \tag{1}$$

and

$$F(L, \langle L \rangle) \propto \exp\left[-\frac{(L-\langle L \rangle)^2}{2h\langle L \rangle}\right], \tag{2}$$

where D and L are the NW diameter and length, $\langle \rangle$ denote the mean values, and *h* and *w* are the ML thicknesses to the NW growth direction (111) and side facet (110) directions, respectively. The Poissonian standard deviations σ_p for the diameter and length distributions are 5.3 nm and 37.5 nm, respectively. The diameter distribution (Fig. 3(c)) is almost Poissonian, whereas the length distribution (Fig3(d)) seems to be sub-Poissonian, thus further demonstrating the homogeneity of the NW length distribution. The inset in Fig 3(d) shows a positive correlation of 0.5 between the NW length and droplet diameters after 60 min NW growth.

This data gives new insight into the NW growth kinetics on lithography-free oxide patterns. When the As cracker valve is opened, the droplets start to approach supersaturation and simultaneously compete for the amount of Ga with other droplets within the Ga diffusion length. At this point, the nearest neighbor distance, which is known to affect the NW diameters in site-selective growth [24], controls the Ga diffusion into the droplets. Simultaneously the Gibbs-Thompson (G-T) effect [23] allows the larger droplets to supersaturate and start nucleating at lower As concentrations thus elongating them faster which leads to the initial NW size distribution. Droplets have a tendency to self-equilibrate their size distributions in certain growth conditions. This is because the Ga adatom diffusion length along the NW sidewall starts to limit the growth rate of long NWs faster than for short NWs. This partially cancels the G-T effect, which is observed as the correlation between NW length and catalyst droplet diameter (Fig. 3 (c, d)), and could lead to constant length distributions throughout the whole NW growth as seen in Fig. 2 and 3(d) [24]. However, the self-equilibration process [25] offers only partial description of the growth kinetics of our NWs as it does not explain the finite sidewall growth. The linear growth of the NW length indicates that the growth occurs by random nucleation of monolayers, which is expected to induce Poissonian length distributions [26,27]. However, the length distributions seem to be sub-Poissonian, which has not been demonstrated before either for self-catalyzed or Au-catalyzed NWs. Moreover, demonstrations of even Poissonian length distributions are limited to Au-catalyzed InAs NWs [28,29] making this the first demonstration of sub-Poissonian length distribution in epitaxial III-As NWs. The sub-Poissonian distribution is attributed to self-regulated growth caused by nucleation antibunching where the nucleation of each ML consumes a significant amount of group V species from the droplet inducing a temporal anticorrelation between the nucleation events. This allows growth in self-regulated mode, which is expected to result in sub-Poissonian length distributions. [26,30,31] The constant and sub-Poissonian standard deviation of the NW length distribution indicates that the remarkable uniformity of the GaAs NWs grown on lithography-free Si/SiOx patterns is a result of (i) uniformity of the nucleation sites created by the DE of GaAs nanocrystals, spontaneous oxidation and thermal desorption, (ii) lack of secondary nucleation, and (iii) growth under self-regulation caused by nucleation antibunching.

The crystal structure of NWs was studied by TEM to further understand the growth process. TEM images of a NW with preserved Ga droplets of sample NW1 are shown in Fig. 4. In the bottom part of the NW (Fig. 4(b) and (c)) a short, typically less than 150 nm long section of stacking faults and polytypism is observed. It is followed by a remarkably long, up to 2 µm, section of completely twin-free ZB GaAs (Fig. 4(d)). The phase pure ZB ends to a section with sparse twin planes until the top of the NW (Fig. 4(e)). Furthermore, a short WZ segment is found just under the Ga droplets. The disordered section in the bottom part of the NW is formed during the beginning of the NW growth, which may be related to Ga pre-deposition. This assumption is supported by other reports of NWs with similar base structure and rather long Ga pre-deposition times [19] and the absence of the stacking faults when NWs are grown without Ga pre-deposition, or short, under 5 s Ga pre-deposition is used [17,32]. The 2 µm phase pure ZB segments show the excellent crystal quality of our NWs, even though longer ZB segments can be obtained by growing NWs under conditions which prefer growth of thinner NWs [33]. The twin plane formation after the phase pure ZB can be attributed to lateral NW growth. The contact angle between the Ga droplet and solid NW tip decreases towards 90 degrees and the spreading of the droplet over the edges of the NW tip reduces as the NW diameter grows. This is known to increase probability of nucleation at triple phase line which induces twin planes. Thus more and more twinning is seen in the uppermost part of the NW. [17, 34,35,36] The WZ segment under the catalyst droplet, as the one seen in Fig. 4, is a typical feature in self-catalyzed GaAs NWs. It is associated with the rapid change in growth conditions as the material fluxes are terminated [17, 34,37]. Some As species may still be present in the growth chamber after the Ga shutter has been closed as the As cracker needle valve operation is slower than the Ga shutter operation and group V materials are known to remain longer time periods in the growth chamber. This rapidly increases the local V/III-ratio at the droplet, which induces WZ formation in GaAs NWs [38]. Hence, this WZ segment is probably formed from the remaining As species after shutting the Ga flux.

TEM images of a NW with crystallized Ga catalyst droplet from sample NW2 is shown in Fig. 5. The bottom parts of the NWs with crystallized Ga droplets exhibit similar crystal structure to the ones with preserved Ga droplets, as expected. The NW section formed during the Ga catalyst droplet crystallization begins with a short section of stacking faults Fig. 5(b) and (c) that is followed by a 150 nm long section of pure WZ GaAs (Fig. 5(d) and (e)). A short section ~20 nm of ZB is found in the very end of the NWs (Fig. 5(e)). Similar characteristics are typical for self-catalyzed GaAs NWs with crystallized catalyst droplets [32,37,39]. Based on the TEM-analysis GaAs NWs grown on lithography-free Si/SiOx patterns exhibit typical characteristics of high-quality self-catalyzed GaAs NWs.

HR-XRD was used to obtain statistically meaningful data of the NW crystal structure. HR-XRD spectra of the samples NW1-NW4 are presented in Fig. 6(a) and symmetric reciprocal space maps (RSMs) of samples NW1-NW3 are shown in Fig. 6(b)-(d), respectively. The Si(111) and ZB GaAs(111) reflections are clearly seen in all of the spectra, disregarding spectrum of sample NW4 which has too small NWs to gain adequate intensity from GaAs(111) reflection. The GaAs(111) reflections in Fig. 6(a) show some degree of broadening towards low omega-2theta direction which can be attributed to the presence of short period polytypism [40] at the lower part of the NWs as well as in the upper part of samples NW2 and NW3. In addition to the GaAs(111) reflection, a weak peak can be observed in the smaller omega-2theta side of the GaAs(111) reflection in the spectra of samples NW2 and NW3 (Fig. 6(a)). The existence of this reflection is confirmed by the RSMs of samples with crystallized Ga catalyst droplets (NW2 and NW3 in Fig. 6 (c) and (d)) and it is attributed to WZ GaAs(0002) reflection. Because the phase pure WZ segments in NW sections formed during droplet crystallization are 150 nm long the WZ reflection is assumed to be strain free which enables determination of WZ GaAs lattice constant $c$. We have obtained a WZ GaAs lattice constant $c$=6.5689 Å by comparing the WZ GaAs(0002) reflection with the Si(111) reflection in RSM in Fig. 6(b) where WZ peak is most pronounced. This value is consistent with WZ GaAs lattice constant measured from Au-catalyzed WZ-GaAs NWs using asymmetric RSMs 6.5701 Å [41]. As a test of the accuracy of the method, we determined also the ZB GaAs lattice constant and obtained value of 5.6524 Å which deviates from the literature value [42] by only 0.015

%. These observations demonstrate that HR-XRD is a valuable tool for NW characterization providing complementary statistical information for TEM analysis performed for a small number of NWs per sample.

Furthermore, the optical properties of the NWs were studied using low-temperature micro-PL in order to correlate them with the crystalline structure. Power-dependent micro-PL spectra of samples NW1-NW3 are shown in Fig. 7. The peak wavelength is between 840 nm and 855 nm, which is significantly redshifted with respect to bulk GaAs band edge emission (816 nm) at 10 K. The spectra are broadened to longer wavelength side and the amount of fine structure in the spectra seems to increase as the NWs get smaller (from Fig. 6(a)-(c)). Possible reasons for the redshift are unintentional Si doping solute into the Ga catalyst droplet from the substrate [43], carbon or other impurity [44] or band bending due to NW surface oxidation [45,46]. The fine structure emission is most probably related to indirect and localized transitions at sections with polytypism and stacking faults [47,48]. The spectra obtained from sample NW1 shows least fine structure as it consists mostly of defect free ZB except for the short disordered section at the bottom of the NW and the few twin planes in the upper part of the NWs. Furthermore, samples NW2 and NW3 have equally long polytypic sections at their bottom and top parts but sample NW2 has 1.5 µm longer pure ZB section. Hence, the fine structure emission becomes more prominent when the relative amount of stacking faults increases in shorter NWs, or in other words, the pure ZB segments become shorter in relation to the charge carrier diffusion length. Thus, we are able to associate the characteristics of the micro-PL spectra to the crystal structure of the NWs.

**Conclusions**

Based on the structural and optical analysis presented in this work, we have shown that the lithography-free oxide patterns provide a suitable template for growth of high-quality GaAs NWs by Ga-catalyzed technique. The sub-Poissonian length distribution of our NWs is a demonstration of their remarkable uniformity, which is attributed to the uniformity of the nucleation sites, lack of secondary nucleation, and self-regulated growth mode. However, further experiments are required for forming a complete understanding of the sub-Poissonian nature of the GaAs NWs grown on lithography-free oxide patterns. Furthermore, the HR-TEM and HR-XRD studies revealed that the GaAs NWs grown on lithography-free oxide patterns exhibit high crystalline quality characterized by up to 2 μm long defect free ZB segments. The WZ segments formed during Ga catalyst droplet crystallization were used for the determination of lattice constant $c$=6.5689 Å which corroborates the previously presented value obtained for phase pure WZ NWs. The micro-PL results correlated with the NW crystal structure so that the amount of fine structure PL emission relates to the proportion of disordered sections in the NWs. More generally speaking, our observations show that the non-destructive micro-PL and HR-XRD techniques are valuable tools for the characterization of NW crystal structure and, in particular, provide complementary statistically significant data to support the TEM results.

**Acknowledgements**

This work made use of Aalto University Nanomicroscopy Center (Aalto-NMC) facilities. Funding from the Academy of Finland Project NESP (decision number 294630) is acknowledged. We thank Dr. Soile Suomalainen for critical reading of the manuscript.

**References**

[1] Glas F. Critical dimensions for the plastic relaxation of strained axial heterostructures in free-standing nanowires. Phys. Rev. B 2006;74:12.

[2] Mayer B, Janker L, Loitsch B, Treu J, Kostenbader T, Lichtmannecker S, Reichert T, Morkötter S, Kaniber M, Abstreiter G, Gies C, Koblmüller G, Finley JJ. Monolithically Integrated High-β Nanowire Lasers on Silicon. Nano Lett. 2016;16:1.

[3] Couteau C., Larrue A., Wilhelm C., Soci C. Nanowire Lasers. Nanophotonics 2015;4:.

[4] Tomioka K, Motohisa J, Hara S, Hiruma K, Fukui T. GaAs/AlGaAs Core Multishell Nanowire-Based Light-Emitting Diodes on Si. Nano Lett. 2010;10:5.

[5] Lysov A, Offer M, Gutsche C, Regolin I, Topaloglu S, Geller M, Prost W, Tegude FJ. Optical properties of heavily doped GaAs nanowires and electroluminescent nanowire structures. Nanotechnology 2011;22:8.

[6] Wallentin J, Anttu N, Asoli D, Huffman M, Åberg I, Magnusson MH, Siefer G, Fuss-Kailuweit P, Dimroth F, Witzigmann B, Xu HQ, Samuelson L, Deppert K, Borgström MT. InP Nanowire Array Solar Cells Achieving 13.8% Efficiency by Exceeding the Ray Optics Limit. Science 2013;339:6123.

[7] Holm JV, Jørgensen HI, Krogstrup P, Nygård J, Liu H, Aagesen M. Surface-passivated GaAsP single-nanowire solar cells exceeding 10% efficiency grown on silicon. Nat Commun 2013;4:.

[8] Tomioka K, Yoshimura M, Fukui T. A III-V nanowire channel on silicon for high-performance vertical transistors. Nature 2012;488:7410.

[9] Dayeh S, Aplin DP, Zhou X, Yu PK, Yu E, Wang D. High Electron Mobility InAs Nanowire Field-Effect Transistors. Small 2007;3:2.

[10] Reimer ME, Bulgarini G, Akopian N, Hocevar M, Bavinck MB, Verheijen MA, Bakkers EPAM, Kouwenhoven LP, Zwiller V. Bright single-photon sources in bottom-up tailored nanowires. Nat Commun 2012;3:.

[11] Holmes MJ, Choi K, Kako S, Arita M, Arakawa Y. Room-Temperature Triggered Single Photon Emission from a III-Nitride Site-Controlled Nanowire Quantum Dot. Nano Lett. 2014;14:2.

[12] Spirkoska D, Efros AL, Lambrecht WRL, Cheiwchanchamnangij T, Fontcuberta iM, Abstreiter G. Valence band structure of polytypic zinc-blende/wurtzite GaAs nanowires probed by polarization-dependent photoluminescence. Phys. Rev. B 2012;85:4.

[13] Mukherjee A, Ghosh S, Breuer S, Jahn U, Geelhaar L, Grahn HT. Spatially resolved study of polarized micro-photoluminescence spectroscopy on single GaAs nanowires with mixed zincblende and wurtzite phases. J. Appl. Phys. 2015;117:5.

[14] Heiss M, Russo-Averchi E, Dalmau-Mallorquí A, Tütüncüoğlu G, Matteini F, Rüffer D, Conesa-Boj S, Demichel O, Alarcon-Lladó E, Fontcuberta i Morral A. III-V nanowire arrays: growth and light interaction. Nanotechnology 2014;25:1.

[15] Larrue A, Wilhelm C, Vest G, Combrié S, De Rossi A, Soci C. Monolithic integration of III-V nanowire with photonic crystal microcavity for vertical light emission. Opt. Express 2012;20:7.

[16] Colombo C, Spirkoska D, Frimmer M, Abstreiter G, Fontcuberta iM. Ga-assisted catalyst-free growth mechanism of GaAs nanowires by molecular beam epitaxy. Phys. Rev. B 2008;77:15.

[17] Cirlin GE, Dubrovskii VG, Samsonenko YB, Bouravleuv AD, Durose K, Proskuryakov YY, Mendes B, Bowen L, Kaliteevski MA, Abram RA, Zeze D. Self-catalyzed, pure zincblende GaAs nanowires grown on Si(111) by molecular beam epitaxy. Phys. Rev. B 2010;82:3.

[18] Matteini F, Tütüncüoğlu G, Rüffer D, Alarcón-Lladó E, Fontcuberta i Morral A. Ga-assisted growth of GaAs nanowires on silicon, comparison of surface SiOx of different nature. J. Cryst. Growth 2014;404:.

[19] Munshi AM, Dheeraj DL, Fauske VT, Kim DC, Huh J, Reinertsen JF, Ahtapodov L, Lee KD, Heidari B, van Helvoort, A. T. J., Fimland BO, Weman H. Position-Controlled Uniform GaAs Nanowires on Silicon using Nanoimprint Lithography. Nano Lett. 2014;14:2.

[20] Plissard S, Larrieu G, Wallart X, Caroff P. High yield of self-catalyzed GaAs nanowire arrays grown on silicon via gallium droplet positioning. Nanotechnology 2011;22:27.

[21] Hakkarainen TV, Schramm A, Mäkelä J, Laukkanen P, Guina M. Lithography-free oxide patterns as templates for self-catalyzed growth of highly uniform GaAs nanowires on Si(111). Nanotechnology 2015;26:27.

[22] Rudolph D, Schweickert L, Morkötter S, Loitsch B, Hertenberger S, Becker J, Bichler M, Abstreiter G, Finley JJ, Koblmüller G. Effect of interwire separation on growth kinetics and properties of site-selective GaAs nanowires. Appl. Phys. Lett. 2014;105:3.

[23] Zhang Y, Sanchez AM, Sun Y, Wu J, Aagesen M, Huo S, Kim D, Jurczak P, Xu X, Liu H. Influence of Droplet Size on the Growth of Self-Catalyzed Ternary GaAsP Nanowires. Nano Lett. 2016;16:2.

[24] Dubrovskii VG, Xu T, Álvarez AD, Plissard SR, Caroff P, Glas F, Grandidier B. Self-Equilibration of the Diameter of Ga-Catalyzed GaAs Nanowires. Nano Lett. 2015;15:8.

[25] Tersoff J. Stable Self-Catalyzed Growth of III-V Nanowires. Nano Lett. 2015;15:10.

[26] Dubrovskii VG. Self-regulated pulsed nucleation in catalyzed nanowire growth. Phys. Rev. B 2013;87:19.

[27] Dubrovskii VG. Kinetic narrowing of size distribution. Phys. Rev. B 2016;93:17.

[28] Dubrovskii VG, Sibirev NV, Berdnikov Y, Gomes UP, Ercolani D, Zannier V, Sorba L. Length distributions of Au-catalyzed and In-catalyzed InAs nanowires. Nanotechnology 2016;27:37.

[29] Dubrovskii VG, Berdnikov Y, Schmidtbauer J, Borg M, Storm K, Deppert K, Johansson J. Length Distributions of Nanowires Growing by Surface Diffusion. Crystal Growth & Design 2016;16:4.

[30] Glas F, Harmand J, Patriarche G. Nucleation Antibunching in Catalyst-Assisted Nanowire Growth. Phys. Rev. Lett. 2010;104:13.

[31] Glas F. Statistics of sub-Poissonian nucleation in a nanophase. Phys. Rev. B 2014;90:12.

[32] Plissard S, Dick KA, Larrieu Guilhem, Godey Sylvie, Addad Ahmed, Wallart X, Caroff Philippe. Gold-free growth of GaAs nanowires on silicon: arrays and polytypism. Nanotechnology 2010;21:38.

[33] Krogstrup P, Popovitz-Biro R, Johnson E, Madsen MH, Nygård J, Shtrikman H. Structural Phase Control in Self-Catalyzed Growth of GaAs Nanowires on Silicon (111). Nano Lett. 2010;10:11.

[34] Bauer B, Rudolph A, Soda M, Fontcuberta i Morral A, Zweck J, Schuh D, Reiger E. Position controlled self-catalyzed growth of GaAs nanowires by molecular beam epitaxy. Nanotechnology 2010;21:43.

[35] Glas F, Harmand J, Patriarche G. Why Does Wurtzite Form in Nanowires of III-V Zinc Blende Semiconductors? Phys. Rev. Lett. 2007;99:14.

[36] Krogstrup P, Curiotto S, Johnson E, Aagesen M, Nygård J, Chatain D. Impact of the Liquid Phase Shape on the Structure of III-V Nanowires. Phys. Rev. Lett. 2011;106:12.

[37] Yu X, Wang H, Lu J, Zhao J, Misuraca J, Xiong P, von Molnár S. Evidence for Structural Phase Transitions Induced by the Triple Phase Line Shift in Self-Catalyzed GaAs Nanowires. Nano Lett. 2012;12:10.

[38] Krogstrup P, Hannibal Madsen M, Hu W, Kozu M, Nakata Y, Nygård J, Takahasi M, Feidenhans'l R. In-situ x-ray characterization of wurtzite formation in GaAs nanowires. Appl. Phys. Lett. 2012;100:9.

[39] Heon Kim Y, Woo Park D, Jun Lee S. Gallium-droplet behaviors of self-catalyzed GaAs nanowires: A transmission electron microscopy study. Appl. Phys. Lett. 2012;100:3.

[40] Schroth P, Köhl M, Hornung J, Dimakis E, Somaschini C, Geelhaar L, Biermanns A, Bauer S, Lazarev S, Pietsch U, Baumbach T. Evolution of Polytypism in GaAs Nanowires during Growth Revealed by Time-Resolved *in situ* x-ray Diffraction. Phys. Rev. Lett. 2015;114:5.

[41] Jacobsson D, Yang F, Hillerich K, Lenrick F, Lehmann S, Kriegner D, Stangl J, Wallenberg LR, Dick KA, Johansson J. Phase Transformation in Radially Merged Wurtzite GaAs Nanowires. Crystal Growth & Design 2015;15:10.

[42] Vurgaftman I, Meyer JR, Ram-Mohan LR. Band parameters for III–V compound semiconductors and their alloys. J. Appl. Phys. 2001;89:11.

[43] Keck PH, Broder J. The Solubility of Silicon and Germanium in Gallium and Indium. Phys. Rev. 1953;90:4.

[44] Scolfaro LMR, Pintanel R, Gomes VMS, Leite JR, Chaves AS. Impurity levels induced by a C impurity in GaAs. Phys. Rev. B 1986;34:10.

[45] Speckbacher M, Treu J, Whittles TJ, Linhart WM, Xu X, Saller K, Dhanak VR, Abstreiter G, Finley JJ, Veal TD, Koblmüller G. Direct Measurements of Fermi Level Pinning at the Surface of Intrinsically n-Type InGaAs Nanowires. Nano Lett. 2016;16:8.

[46] Songmuang R, Giang LTT, Bleuse J, Den Hertog M, Niquet YM, Dang LS, Mariette H. Determination of the Optimal Shell Thickness for Self-Catalyzed GaAs/AlGaAs Core-Shell Nanowires on Silicon. Nano Lett. 2016;16:6.

[47] Graham AM, Corfdir P, Heiss M, Conesa-Boj S, Uccelli E, Fontcuberta iM, Phillips RT. Exciton localization mechanisms in wurtzite/zinc-blende GaAs nanowires. Phys. Rev. B 2013;87:12.

[48] Spirkoska D, Arbiol J, Gustafsson A, Conesa-Boj S, Glas F, Zardo I, Heigoldt M, Gass MH, Bleloch AL, Estrade S, Kaniber M, Rossler J, Peiro F, Morante JR, Abstreiter G, Samuelson L, Fontcuberta iM. Structural and optical properties of high quality zinc-blende/wurtzite GaAs nanowire heterostructures. Phys. Rev. B 2009;80:24.

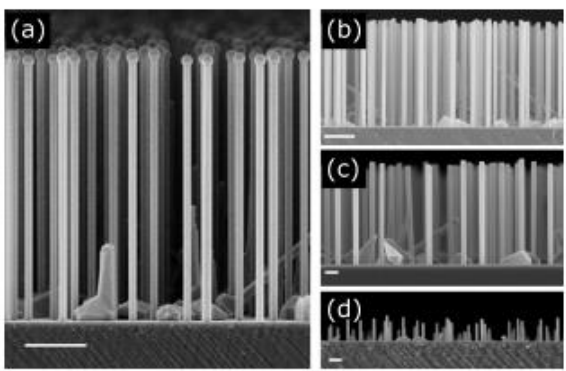

**Figure 1.** SEM pictures of samples NW1-NW4 in (a)-(d), respectively. The scale bars are 1 µm in (a) and (b), and 200 nm in (c) and (d).

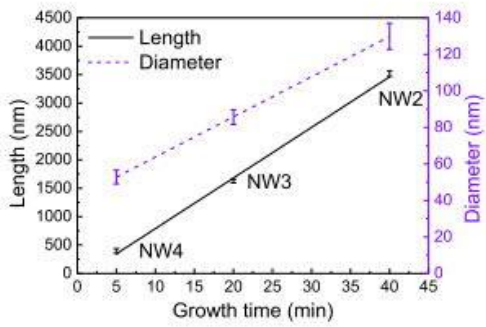

**Figure 2.** NW length (solid line) and diameter (dashed line) data of samples NW2, NW3, and NW4 presented with the standard deviations.

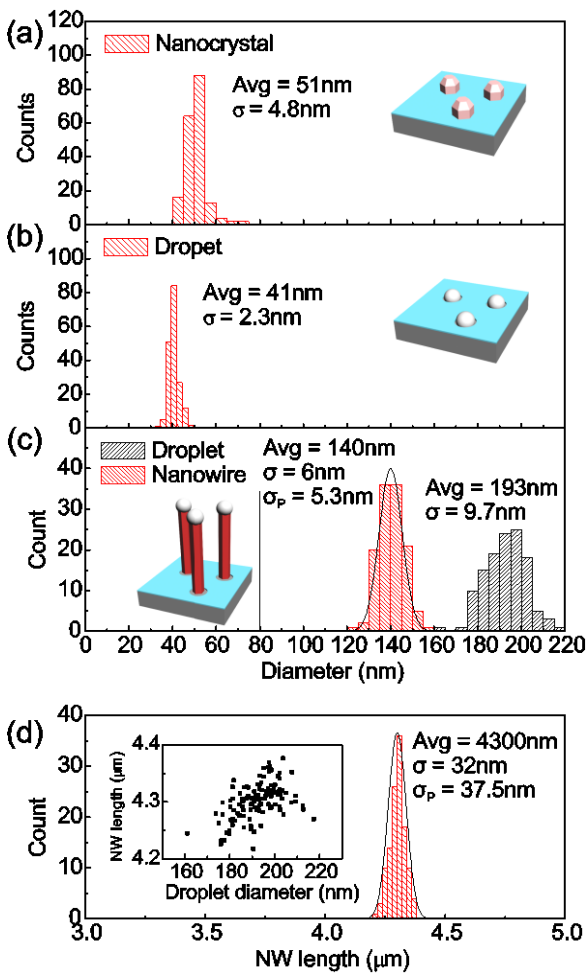

**Figure 3.** Size distributions at different steps of the self-catalyzed GaAs NW growth on lithography-free Si/SiOx patterns. (a) Diameter of the GaAs nanocrystals grown on oxide free Si(111) by droplet epitaxy. (b) Diameter of the Ga catalyst droplets deposited in oxide free areas formed by thermal desorption of the GaAs

nanocrystals. (c) Diameter of the NWs with a Poissonian fit and Ga catalyst droplet diameters after 60 min growth of sample NW1. (d) NW length of sample NW1 with a Poissonian fit after 60 min growth. The measured and calculated Poissonian standard deviations are denoted by σ and $σ_P$, respectively. The inset in Fig. (d) shows the correlation between NW length and Ga catalyst droplet diameter.

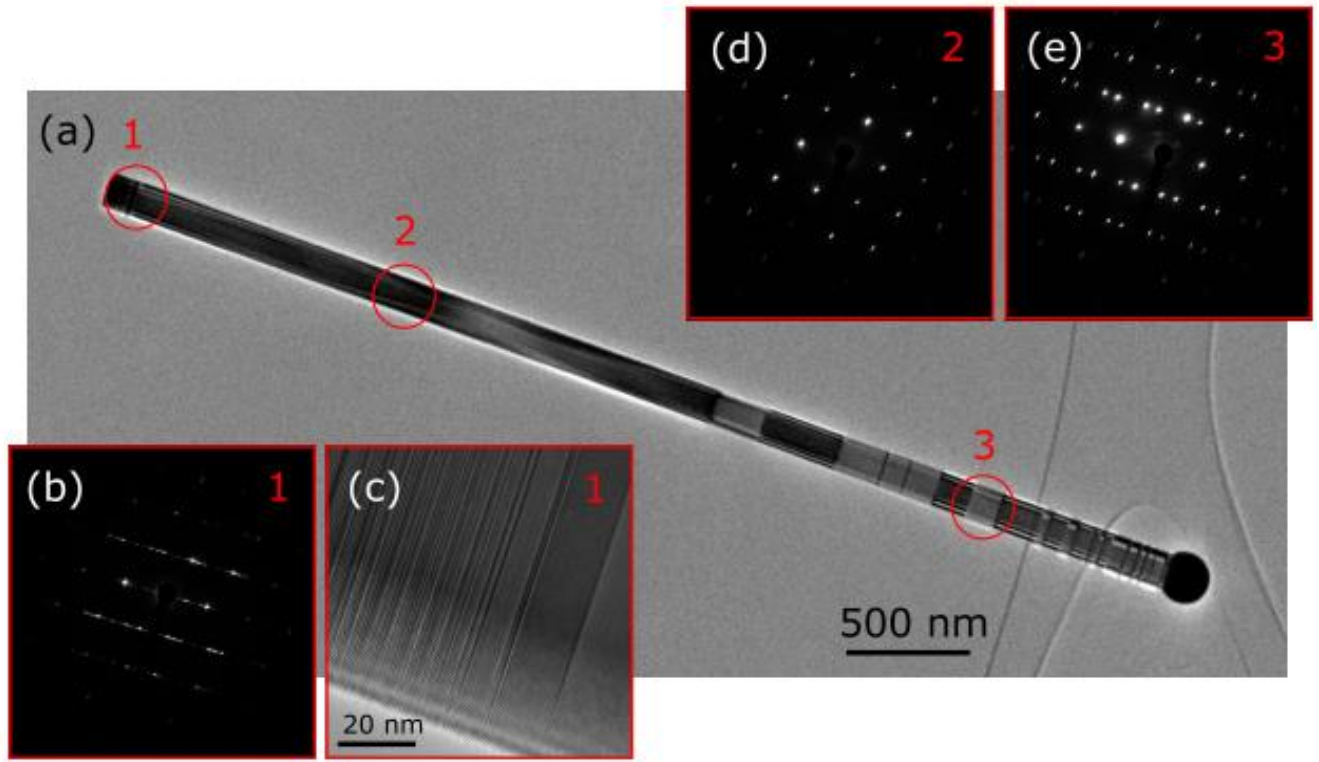

**Figure 4.** TEM images of sample NW1. (a) HR-TEM low magnification overview. (b) and (c) SAED pattern and HR-TEM micrograph of the disordered section at root of the wire, respectively. (d) SAED pattern of the pure ZB section. (e) SAED pattern of the twinned ZB section.

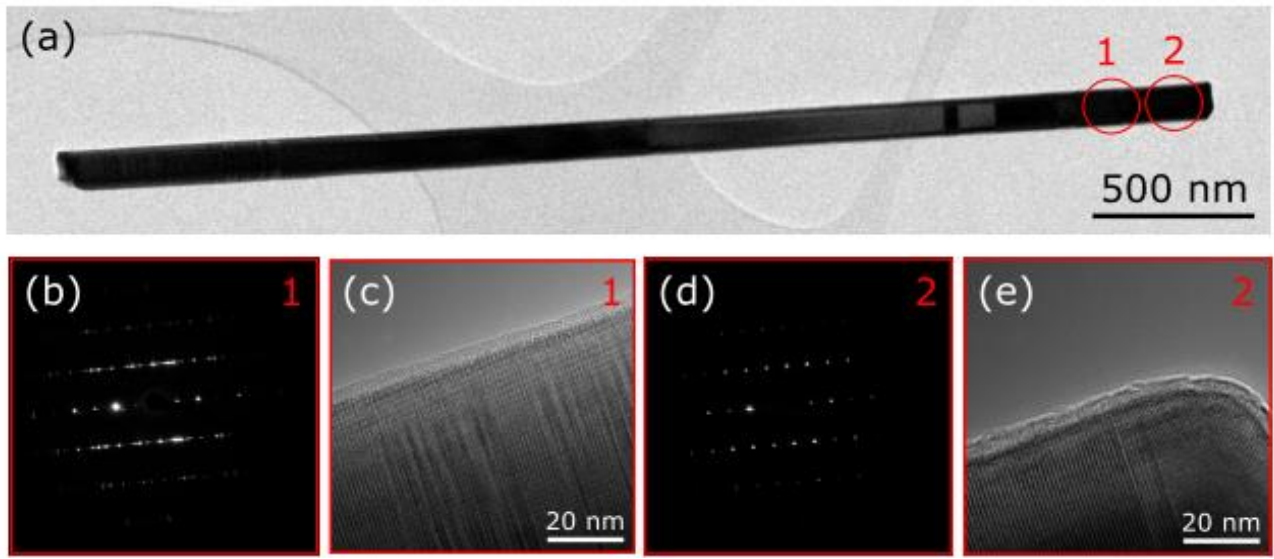

**Figure 5.** TEM images of sample NW2. (a) HR-TEM low magnification overview. (b) SAED patter and HR-TEM micrograph of the disordered section close to tip of the NW, respectively. (c) SAED pattern of the pure WZ section at the tip of the NW. (d) HR-TEM micrograph of the very tip of the NW.

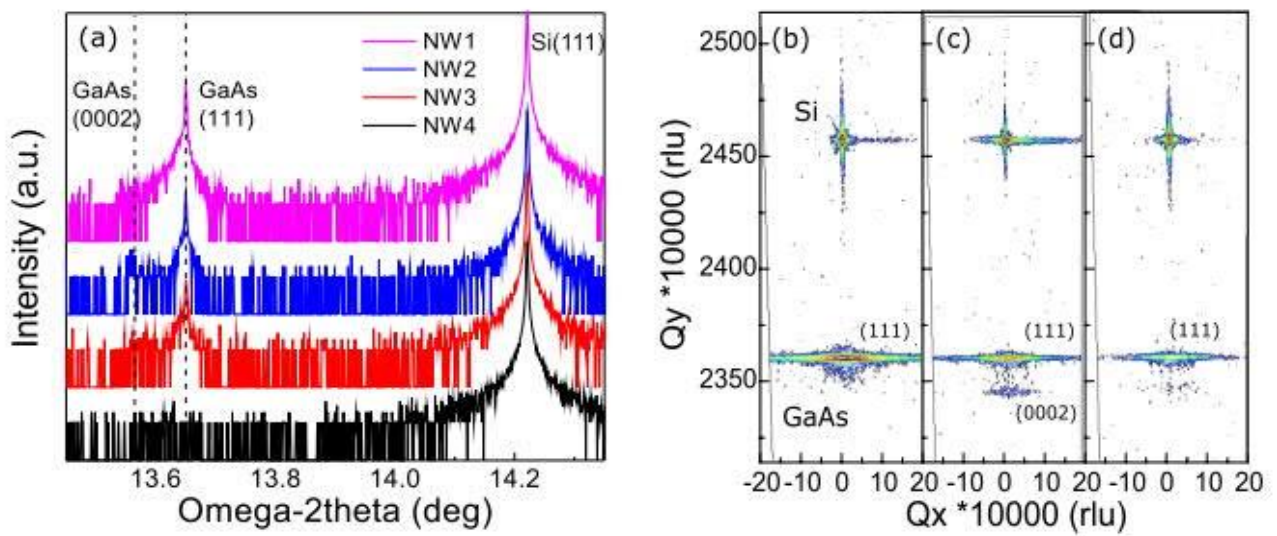

**Figure 6.** HR-XRD spectra of samples NW1-NW4 from top to bottom in (a) and reciprocal space maps of samples NW1-NW3 in (b)-(d), respectively.

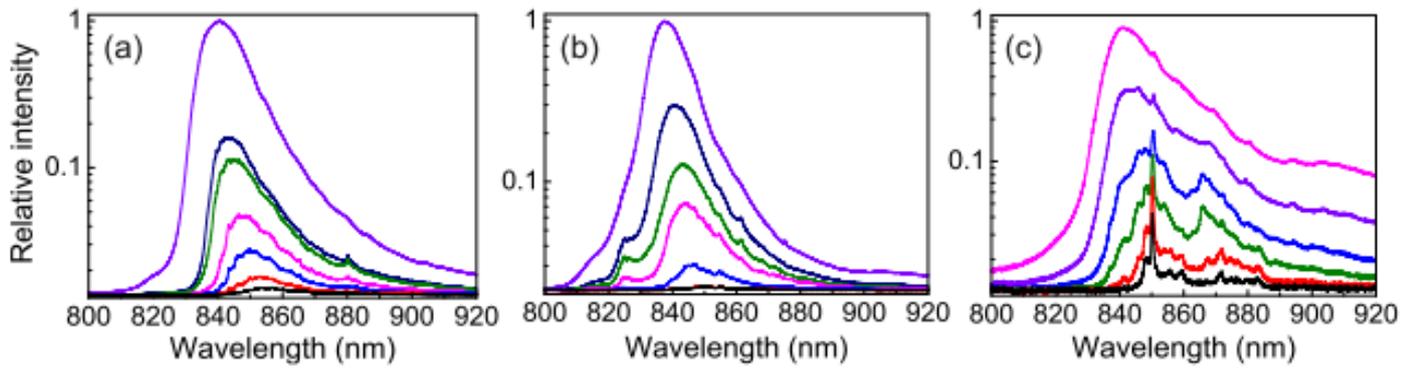

**Figure 7.** Micro-photoluminescence spectra of samples NW1-NW3 in (a)-(c), respectively. The used excitation power densities vary from 0.1 to 96 Wcm$^{-2}$ in (a) and (b), and from 1 to 960 Wcm$^{-2}$.